\documentclass[prd,aps,showpacs,superscriptaddress,floatfix,preprintnumbers,reprint,nofootinbib]{revtex4-2}
\usepackage{hyperref}
\usepackage{graphicx}
\usepackage{siunitx}
\usepackage[english]{babel}

\begin{document}

\title{Preliminary Investigation of a Higgs Factory based on Proton-Driven Plasma Wakefield Acceleration}

\author{J.~Farmer}
\email{j.farmer@cern.ch}
\affiliation{Max-Planck-Institut f\"ur Physik, M\"unchen, Germany}
\author{A.~Caldwell}
\email{caldwell@mpp.mpg.de}
\affiliation{Max-Planck-Institut f\"ur Physik, M\"unchen, Germany}
\author{A.~Pukhov}
\email{pukhov@tp1.uni-duesseldorf.de}
\affiliation{Heinrich-Heine-Universit\"at D\"usseldorf, D\"usseldorf, Germany}

\begin{abstract}
A Higgs Factory is considered the highest priority next collider project by the high-energy physics community.  Very advanced designs based on radio-frequency cavities exist, and variations on this approach are still being developed.  Recently, an option based on electron-bunch driven plasma wakefield acceleration has also been proposed.  In this article, we discuss a further option based on proton-driven plasma wakefield acceleration.  
This option has significant potential advantages due to the high energy of the plasma wakefield driver, simplifying the plasma acceleration stage.  Its success will depend on further developments in producing compact high-energy proton bunches at a high rate, which would also make possible a broad range of synergistic particle-physics research.
\end{abstract}
\maketitle

\section{Introduction}
 With the discovery of the Higgs Boson in 2012~\cite{ref:Higgs1,ref:Higgs2}, attention in the high-energy physics community has turned to particle colliders that would allow its detailed study.  Indeed, several variants of a Higgs Factory are actively discussed today, including linear~\cite{ref:Linear1,ref:Linear2} and circular~\cite{ref:Circular} colliders using RF acceleration to bring electrons and positrons to high energy.  Recently, a scheme based on electron-bunch driven plasma-wakefield acceleration (PWFA)~\cite{ref:HALHF} was also proposed.  In this article, we discuss the possibility of a proton-bunch driven plasma-wakefield acceleration scheme.  We find that such a scheme is an attractive option as 1) it is more compact than the conventional RF acceleration schemes, and 2) is simpler than the electron-driven PWFA scheme as staging is not necessary.  The use of a quasilinear wakefield may also offer advantages for the acceleration of positrons.  Significant challenges are present, however, in providing the desired luminosity. As such, this article provides a conceptual basis for such a facility in the hope that the challenges described can be met.
 
 In general, acceleration in plasma generates interest due to the very strong accelerating gradients it allows. Initially, laser driven plasma wakefield acceleration was considered in the literature~\cite{ref:TajimaDawson} and it was later recognized that the plasma could also be excited by an electron bunch~\cite{ref:Chen}.  More recently, proton-driven plasma wakefield acceleration was also introduced~\cite{ref:Caldwell-protondriven}.
 It has been demonstrated that plasmas can be used for accelerating particles to relativistic energies;
gradients well above $1$~GV/m have been achieved.   For a recent review, see, e.g., reference~\cite{Ferrario:2019pkx}.  
The energy in a short laser pulse or electron bunch is not sufficient to bring bunches of particles to the energies needed for a high-energy collider.  In order to reach high energy, many acceleration stages must therefore be deployed, each using a fresh driver. 
The advantage of a proton driver is that the energy of the driver, using today's technology, is sufficient to easily reach accelerated bunch particle energies required for a Higgs Factory without staging, greatly simplifying the accelerator complex. A limitation of the proton-driven scheme has been the repetition rate of the driver, which limits the achievable luminosity.  With the development of fast-ramping superconducting magnets~\cite{ref:Piekarz,Piekarz:2019yis} based on high-temperature superconductors, this limitation can be largely removed, making the proton-driven scheme very attractive.  The use of high-temperature superconductors would significantly reduce the required cryogenic power.

In this paper, we briefly review the proton-driven plasma acceleration scheme. We then discuss a possible layout for the proton acceleration complex and the proton bunch parameters that would be required for our scheme. This is followed by a discussion of the plasma beamline needs.  We have simulated acceleration in our proton-driven scheme, assuming that a suitable driver would be available.  The simulations demonstrate that the required energy can be achieved for both electrons and positrons, but the control of the witness beam quality is only demonstrated for the electron bunch. Significant further work will be required to investigate the feasibility of accelerating positrons.  We then evaluate the luminosity for a Higgs Factory based on our set of assumptions for the proton-driven acceleration complex and the plasma-based acceleration stage, provided that a suitable solution for positron acceleration can be found.  Final focus parameters are taken from other studies.  We conclude with a discussion of the developments that would be necessary to make this approach a reality.

\section{Proton-driven Plasma Wakefield Acceleration}

Proton-driven plasma wakefield acceleration was initially considered assuming a short proton bunch~\cite{ref:Caldwell-protondriven} as the driver of plasma wakefields.  As short proton bunches are not available today, an R\&D program was developed~\cite{ref:Path} based on the concept of the self-modulation of a long proton bunch~\cite{ref:Modulation}.  The AWAKE Collaboration has carried out a very successful experimental program based on this approach.  A recent summary of AWAKE results can be found here~\cite{ref:AWAKE_future}.  In this paper, we return to the concept of a short proton bunch driver, as it will be more effective for the task at hand.

High-energy proton bunches with $10^{11}$ protons per bunch are readily available today~\cite{ref:LHCinjectors}.  Producing the short bunches assumed for our study is a key element that remains to be demonstrated. We assume a $~400$~\si{\giga\electronvolt}/c proton bunch with a length $\sigma_z\approx 150$~\si{\micro\metre} as our PWFA driver.  
While these parameters are not currently available, we assume a momentum spread of 
10\% to give a longitudinal phase space volume comparable to what can currently be achieved~\cite{ref:AWAKEproposal}.

\subsection{Wakefield structure}

As the short proton bunch travels through plasma, it attracts the plasma electrons and sets them in motion, leading to a plasma wave. For a highly relativistic proton drive bunch, the electric field seen by the plasma electrons is transverse to the direction of the proton beam, and the plasma electrons begin to oscillate with frequency $\omega_p$ given by
$$\omega_p = \sqrt{\frac{n_p e^2}{\epsilon_0 m}},$$
where $n_p$ is the density of plasma electrons.  Given their large mass, the plasma ions move relatively little.  The oscillating electrons initially move toward the proton beam axis, then pass through each other, creating a region of low plasma electron density with very strong electric fields. At fixed plasma density, the pattern moves with the proton bunch velocity. 
In addition to the large accelerating gradients $\sim mc\omega_p/e$, the wakefield structure also provides strong focusing fields.

In order to effectively drive the wakefield, the driver length should be on the order of $c/\omega_p$.  Although lower plasma densities could be used to allow a longer drive bunch, the accelerating gradients are commensurately lower. 

\subsection{Longitudinal Growth and Dephasing}

Dephasing between the electron/positron witness and the proton drive bunches occurs due to their great mass difference.  The electron and positron bunches are effectively moving at the speed of light, while protons with energy $400$~\si{\giga\electronvolt} move slightly below light speed.  This causes the distance between the witness and drive bunches to decrease by a few hundred \si{\micro\metre} after $100$~\si{\metre} of propagation, which corresponds to a large fraction of a wakefield period, causing the witness to move out of the optimal accelerating phase.  
However, although the distance between the drive and witness bunches varies, it is possible to keep the witness in an accelerating phase by controlling the plasma wavelength.  Adjusting the density of the plasma over the acceleration length~\cite{ref:Katsouleas,ref:PukhovDephasing} can compensate for the phase slippage, so that this concern is easily remedied.


The initial momentum spread of the protons will induce some longitudinal growth in the bunch. For a $10$\% momentum spread and a nominal energy of $400$~\si{\giga\electronvolt}, this introduces a growth of about $10$~\si{\micro\metre} in $100$~\si{\metre}.  Of greater concern is the significant loss of energy from protons as they give energy to the plasma.  This correlated energy spread will lead to a much more significant lengthening of the bunch.  This lengthening would ultimately lead to some protons falling into a defocusing field region, and so the acceleration distances will need to be kept rather short.

Based on our simulations below, we anticipate satisfying all these considerations with a nominal proton bunch energy of $400$~\si{\giga\electronvolt} and average accelerating gradients of approximately $0.5$~GV/m. This is roughly $14$ times the average acceleration gradient foreseen for the ILC and $7$ times the foreseen CLIC gradient.

\begin{figure*}[htb] 
    \centering
\includegraphics[width=0.85\textwidth]{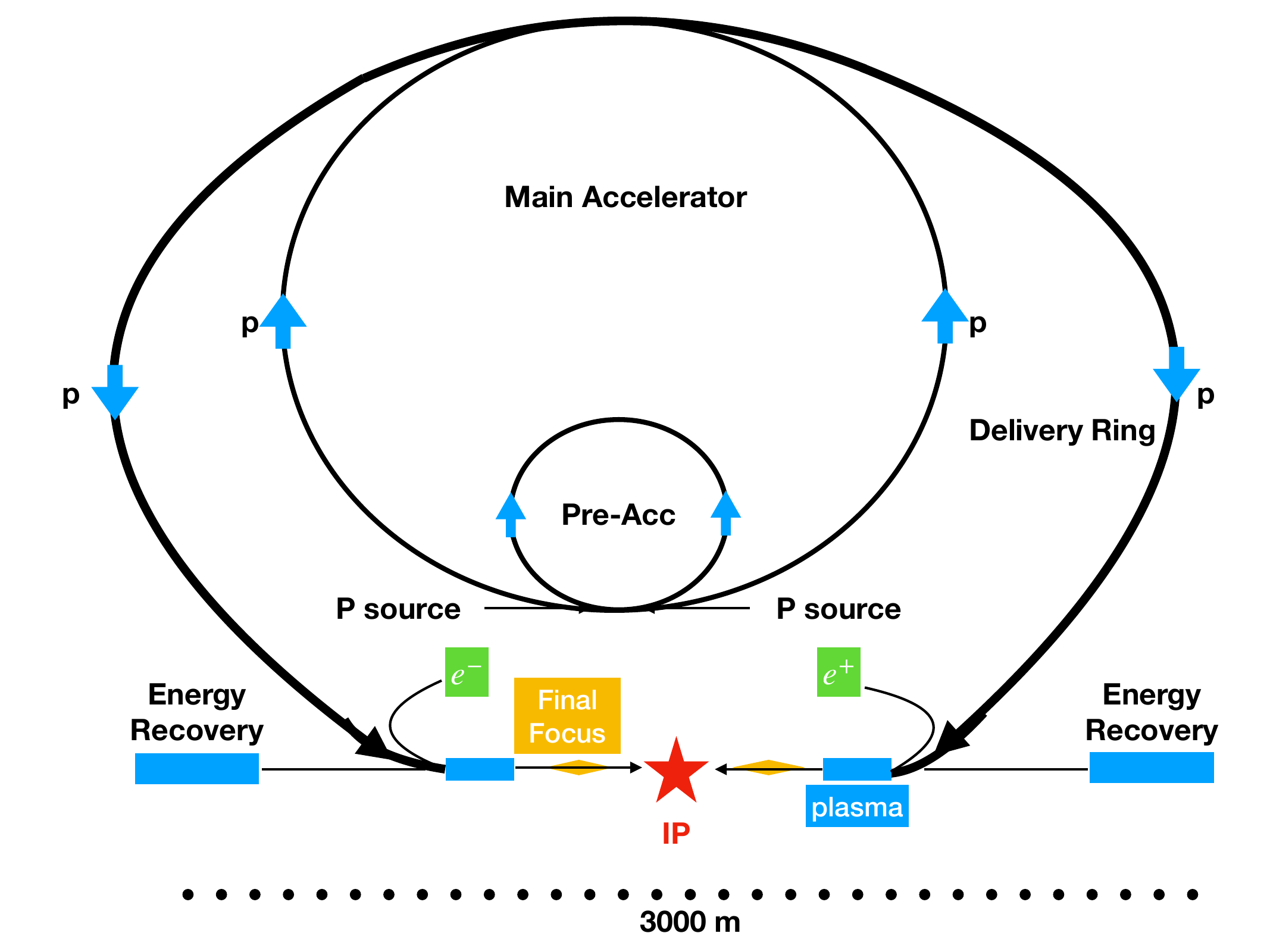}
\caption{\it Schematic accelerator layout for our concept.  A fast-cycling proton synchrotron, fed by a series of pre-accelerators, provides the drive for the plasma wakefields.  The radius of the largest proton accelerator is approximately $1000$~\si{\metre}, while the plasma acceleration stages are about $150$~\si{\metre} long. A final focusing region is indicated, as well as energy recovery plasma sections.}
\label{fig:setup} 
\end{figure*}

\section{Conceptual design for a Higgs factory}
Our scheme for a proton-driven PWFA Higgs Factory is shown in Fig.~\ref{fig:setup}.  A train of proton bunches are brought to the necessary energy in a rapid-cycling synchrotron and then extracted one bunch at a time.  These are then injected into the plasma acceleration sections, trailed by the electron and positron bunches to be accelerated.  The accelerated bunches are then extracted and brought to the interaction point.

The principal advantage of proton bunches as drivers of plasma wakefields, as opposed to lasers or bunches of electrons, is the large energy per particle and per bunch possible with proton beams, thus making proton-driven plasma-based acceleration of charged particles to high energies possible in a single plasma stage.  Furthermore, the quasilinear wakefield structure offers a larger focussing region for positrons, although significant work is required to realise such a concept.  However, today's proton accelerators are not feasible as drivers for a high-energy electron-positron collider as the repetition rate would be too low to produce interesting luminosities.  A further issue is that currently available proton bunches are long, making them unsuitable to drive strong wakefields unless modulated, as done in the AWAKE program~\cite{ref:Karl,ref:Nature,ref:Fabian}. Accelerating electrons and positrons to the required energies for a Higgs Factory has been shown to be feasible in the AWAKE scheme~\cite{ref:Lotov-200}.  However, the energy transfer efficiency based on the self-modulated long proton bunch scheme is expected to be low ($<1$\%).  For the studies presented here, we assume that short ($\sigma_z \approx 150$~\si{\micro\metre}) proton bunches will be available, allowing for higher energy transfer efficiency.

Recently, the push to develop high temperature superconductors (HTS) has opened a possible resolution to the severe repetition rate limitation for proton drivers. Indeed, concepts for synchrotrons based on fast-cycling accelerator magnets capable to operate at $100$'s of T/s ramping rate and reaching field strengths swings of $2$~T have been put forward~\cite{ref:Piekarz} which would allow for significant luminosities in a proton-driven plasma wakefield based Higgs Factory.  In the following, we assume that developments in fast-ramping synchrotrons based on HTS magnet technology continue, and that short proton bunches with $\sigma_z=150$~\si{\micro\metre} will be available at high energy.

\subsection{Proton Accelerator Complex}
As a guide to what may become possible, we consider the Dual Super-Ferric Main Ring (DFSMR) accelerator proposed in \cite{ref:Piekarz2} as baseline.  The DFSMR concept, conceived as based on the FNAL accelerator complex, would allow for two beams of protons, each consisting of 1000 bunches with $10^{11}$ protons per bunch, to be accelerated within a fraction of a second to energies of (in our application) $400$~\si{\giga\electronvolt}.  A $300$~T/s ramping rate has been measured~\cite{Piekarz:2019yis} for HTS-based magnets, while an effective field strength of about $1.3$~T would be needed for a ring with the circumference of the Main Injector at FNAL or SPS at CERN.  We take $0.2$~s for our calculations to accelerate the $1000$ proton bunches to $400$~\si{\giga\electronvolt} and cycle the magnetic fields.  Accelerating gradients capable of giving the protons an energy gain of approximately $80$~\si{\mega\electronvolt}/turn will be required.  The pre-accelerators in the FNAL case consist of a Linac, a Booster and the Main Injector, and they would be tasked with generating 1000 bunches 
with a frequency of $5$~\si{\hertz}.  

The delivery ring is used to bring proton bunches to the plasma acceleration sections.  We assume that one proton bunch is extracted per revolution (period about 20~\si{\micro\second}), giving the plasma time to recover, and therefore one burst consists of 1000 $e^-e^+$ collisions in $20$~\si{\milli\second}, repeating every $200$~\si{\milli\second}.  A further challenge will be the development of fast kicker magnets to extract the single bunches from the ring.  By extracting the last bunch in the train every cycle, a fast rise time of the kicker is necessary and a slower decay/reset time can be accommodated. Rise times on the scale of 10~\si{\nano\second}~\cite{ref:Barnes_kicker} have been already been achieved, but a demonstration for our needs is not available.

The energy contained in each of the two proton beams is 6.4~\si{\mega\joule}, and this would need to be replenished at a rate of $5$~\si{\hertz}, leading to a proton beam power requirement of  $64$~\si{\mega\watt}.  With a wall-plug efficiency of 50\%, a power requirement for accelerating the two beams of proton driver bunches would be approximately 130~\si{\mega\watt}.  In addition, cryogenic power and additional site power would be needed.  Studies on improving klystron efficiency~\cite{ref:cai-klystron} and magnet power needs~\cite{ref:Piekarz} are currently pursued.  
Energy recovery of the spent proton bunches would obviously be beneficial.  First ideas on how energy recovery in plasma wakefield accelerators could be realized have been discussed~\cite{ref:EnergyRecovery1,ref:EnergyRecovery2,ref:EnergyRecovery3}.  Energy recovery would be needed to keep the power needs at levels at or below $150$~\si{\mega\watt}.  These are initial rough estimates that will clearly need more detailed investigations.

\subsection{Proton Bunch Compression}
The bunches accelerated today in proton synchrotrons are quite long, with $\sigma_z \approx 10$~\si{\centi\metre}~\cite{ref:AWAKEproposal}.  However, the momentum spread is small, of order $\frac{\sigma_p}{p} = 10^{-4}$, so that a rotation in phase space would in principle allow the production of proton bunches with $10$\% momentum spread and bunch length of $100$~\si{\micro\metre}.  
A first attempt at designing such a proton bunch compressor is discussed in~\cite{ref:Guoxing}. 
The equilibrium bunch length in a proton storage ring has a quartic-root dependence on the RF voltage and the momentum compaction factor, and so adiabatically varying these quantities within the ring is not a suitable technique to achieve the short bunch lengths required.  A linear chirping approach with a 10\% momentum spread followed by a chicane implies an imposed energy spread approaching $100$~\si{\giga\electronvolt}.  This is comparable to the acceleration required for the electron and positron beams in a Higgs Factory, and so no advantage would be gained from using proton-driven PWFA.


As our scheme relies on the rapid acceleration and extraction of proton bunches, the required stability is lower than for circular colliders.  Acceleration of a pre-compressed bunch in an unstable mode is therefore one possibility to achieve the required bunch length, in which case the energy spread would presumably be lower.  Exploiting transition jumps~\cite{ref:quadrupolepumping} to compress the beam is also a possibility.  Further investigation of these techniques and others is required if the short bunches considered here are to be achieved.  We assume in our simulations that short proton bunches will indeed be available, but allow for a large initial momentum spread of 10\% to have a longitudinal phase-space volume similar to what is available today.  As discussed above, the initial momentum spread is not the limiting factor on the acceleration length, as the longitudinal dispersion due to the driver depositing energy in the plasma is significantly larger.

\subsection{Proton bunch transverse emittance}
The transverse emittance of the proton driver is also important to consider, in order to ensure stable wakefields over long distances. 
As discussed in the simulation section below, we will require sculpted transverse emittances for the proton bunches, with values ranging from $2-58$~\si{\micro\metre}.  This is a further challenge to be met by the proton acceleration complex.  However, we note that the emittance profile used in simulations increases monotonically along the length of the drive bunch, suggesting that is may be possible to harness plasma instabilities to achieve the required tailoring.  The exploitation and control of plasma instabilities has been demonstrated in the AWAKE experiment, where it is used to tailor the drive beam radius~\cite{ref:Karl}.  In a similar way, it may be possible to utilise a short, high-density plasma stage to tailor the beam emittance, although the feasibility of such a technique would need to be investigated.

\subsection{Witness bunch generation and final focus}
The electron and positron bunches required for the scheme are similar to those proposed for injection into the main linac for CLIC~\cite{ref:CLIC_CDR}, and would therefore likely require similar damping rings and compression to reach the required emittance and bunch length.  While not shown in Fig.~\ref{fig:setup}, the associated infrastructure would easily fit within the footprint of the proposed accelerator site.  The required power consumption would be greatly reduced by the development of suitable permanent magnet dipoles~\cite{ref:permanentmagnets}.  Further work is required to investigate the feasibility of shaping the current profile of such beams in order to minimize their energy spread after acceleration.

One important consideration is that damping rings will result in flat beams, i.e.\ beams with a large ratio between the horizontal and vertical emittances, and so a transverse emittance exchange will be necessary to produce round beams for the plasma acceleration stage.  While such emittance exchange has been considered~\cite{ref:Kuske_exchange}, no detailed studies have been performed for our requirements and this will be the subject of further design work.

A similar scheme to increase the aspect ratio after acceleration may better facilitate the manipulation of the beams to maximize the luminosity at the interaction point.  We assume that final focus parameters similar to those discussed, e.g.\ for the ILC~\cite{ref:ILCpars}, will also be possible in our scheme.  

\subsection{Plasma Cell Requirements}
The plasma parameters must be matched to the proton bunch parameters.  For the foreseen proton bunch RMS length of $150$~\si{\micro\metre} and an RMS radius of $240$~\si{\micro\metre}, a plasma density of $\sim3\times10^{14}$~cm$^{-3}$ over a distance of $240$~\si{\metre} is required.  The required density scales as the inverse square of the proton bunch length.

A plasma of this density based on laser-ionized rubidium vapor has been used with great success in the AWAKE experiments~\cite{ref:Plasma} for lengths up to $10$~\si{\metre}, with the same setup planned to be used for total plasma length $20$~\si{\metre} in future runs~\cite{ref:AWAKE_future}. 
A discharge plasma using different noble gases~\cite{ref:Discharge} has also been demonstrated in AWAKE~\cite{ref:AWAKE_filamentation} with comparable densities and a plasma length up to $10$~\si{\metre}.  By utilizing a series of discharges, this technology is expected to scale up to long distances without gaps in the plasma.  

Further development of the discharge technology is required to allow the high repetition rates envisaged in this scheme.  100\% ionization would also be required to avoid field ionization of neutral atoms by the witness bunch, which would act to spoil witness the emittance.
As the proposed acceleration scheme uses a compact driver, the plasma uniformity requirements are relatively mild compared to AWAKE, where resonance between a long train of microbunches must be maintained. We further note that for the short driver, a lighter atom such as lithium could be used to minimize the effect of emittance growth from scattering processes.  The use of lithium would also avoid the potential issue of secondary ionization.

The 20~\si{\micro\second} delay between proton bunch extractions is expected to be sufficient for the plasma to neutralize in the laser ionized vapor case~\cite{ref:PlasmaRecovery2}.  For the discharge source, the 20~\si{\micro\second} delay should be sufficient for the plasma to recover~\cite{ref:PlasmaRecovery1}. Cycling times for discharge plasmas and the required cooling are currently under study~\cite{ref:dArcy}. 

A plasma density gradient will be necessary in order to keep the electron and positron bunches in the optimal accelerating phase.  The gradient acts to compensate for the different velocities of the drive and witness bunches.  The required gradients are moderate (see below) and should be achievable by varying the density of neutral gas before ionization.  This could be achieved through temperature gradients, as in AWAKE~\cite{ref:Karl_gradient}, through the choice of the pumping scheme, or by introducing a series of pressure windows between discharge units.

Assuming the witness bunches absorb 40\% of the energy transferred by the proton bunches to the plasma, we estimate that $\sim12$~kW/m of cooling power will be needed for the plasma device. This could be reduced by adding a third bunch, trailing the witness, to ``mop up'' some fraction of the remaining wakefield energy~\cite{ref:EnergyRecovery3}.  This method has been demonstrated for the case of two laser pulses \cite{ref:cowley_multipulse}.  The requirements on the energy spread and emittance for this third bunch would be significantly more relaxed, and could be used as a further means of energy recovery.

\section{Simulation of electron and positron acceleration}
Proof-of-principle simulations were carried out using the fully 3D quasistatic particle-in-cell code \texttt{qv3d} \cite{pukhov-quasistatic}, built on the \texttt{VLPL} platform \cite{pukhov-vlpl}.

\begin{figure*}[htb] 
    \centering
\includegraphics[width=0.75\textwidth]{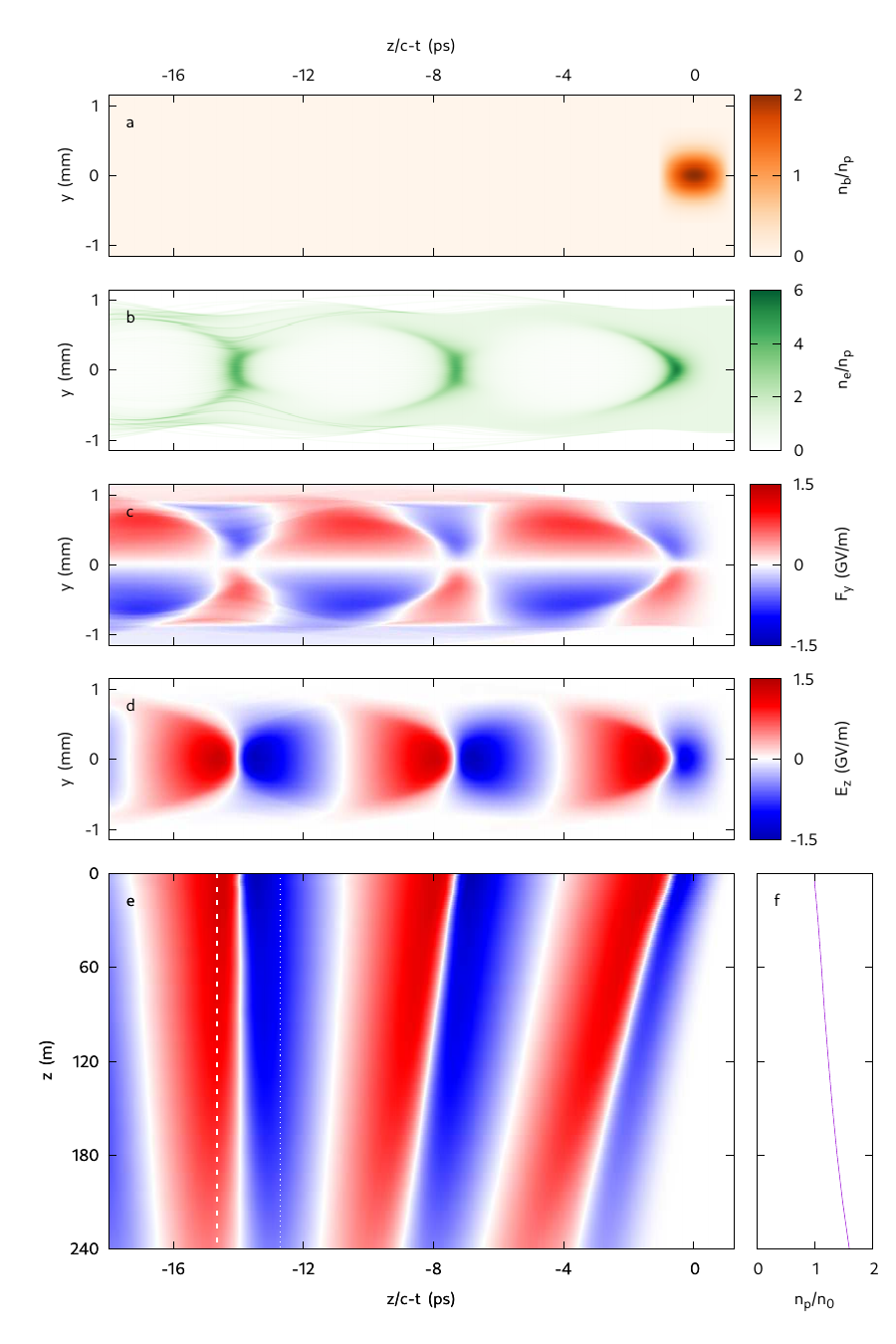}
\caption{\it Overview of the acceleration configuration.  A short proton drive bunch (a) excites a plasma density wave (b), resulting in periodic transverse (c) and longitudinal (d) fields, which can be harnessed to accelerate a witness bunch of electrons or positrons.  The use of a tailored plasma density profile (f) allows the creation of a point of fixed phase in the on-axis longitudinal field (e) over long propagation distances. The injection positions of the electron and positron bunches are indicated by the dotted and dashed lines, respectively. Here, $n_0=n_p(z=0)=3.2\times10^{14}$~\si{\centi\metre^{-3}}.}
\label{fig:phase} 
\end{figure*}

The drive bunch was modeled as a 3D ellipsoid, with a parabolic profile in $z$ and $p_z$, and a Gaussian profile in the transverse planes.  In order to efficiently drive the wakefield, a proton driver with a peak density $2n_p$, an RMS length of $0.5/k_p$ and an RMS radius of $0.8/k_p$ is used.  Choosing the driver population as $1\times10^{11}$, this corresponds to plasma density of $3.2\times10^{14}$~cm$^{-3}$, a skin depth of $1/k_p=300$~\si{\micro\metre}, and an RMS bunch length and radius of 150 and 240~\si{\micro\metre}, respectively.  This driver, shown in Fig.~\ref{fig:phase}a, perturbs the plasma electron density, as shown in Fig.~\ref{fig:phase}b.  The charge separation between plasma electrons and ions leads to periodic transverse and longitudinal fields, shown in Figs.~\ref{fig:phase}c and d.

In order to harness these fields for acceleration, they must remain stable over long distances.  A driver energy of 400~\si{\giga\electronvolt} with an energy spread of 10\% was chosen.  In addition to providing the fields for acceleration of a witness bunch, the plasma wakefields will act on the driver itself, increasing the initial energy spread and acting to limit the acceleration distance. To prevent the driver from rapidly pinching due to the focusing forces, a tailored emittance profile is used, with a constant normalized emittance of 2~\si{\micro\metre} at the bunch head, which then rises linearly from a point 270~\si{\micro\metre} ahead of the bunch center, reaching 58~\si{\micro\metre} at the beam tail.

Combining these techniques allows stable wakefields to be generated over long acceleration distances, as shown in Fig.~\ref{fig:phase}e, which shows the on-axis accelerating field over 240~\si{\metre} of propagation.  A tailored plasma density profile, shown in Fig.~\ref{fig:phase}f, is used to modify the plasma wakelength over the propagation distance.  This is chosen to yield a point of constant phase in the wakefields 14~ps behind the initial driver centroid.  Using the same driver configuration for both electrons and positrons, electrons would be placed ahead of, and positrons behind, this point, indicated by the vertical dotted and dashed lines in the plot.

The wakefields gradually decease over the long propagation distances considered here.  This is partly due to the initial divergence of the beam which acts to ``erode'' the driver, as the wakefields near the head of the bunch are not sufficiently strong to trap the protons.  Dispersion also plays a role, with the length of the driver increasing as protons lose energy to the plasma, reducing the effectiveness with which the bunch drives the wake.  The divergence of the beam head could potentially be reduced by using a driver with a tailored current profile, with a low-current pedestal acting to guide the main bunch.  External focusing fields could also be used to prevent the loss of particles from the driver~\cite{ref:Nature}.  However, the erosion of the beam head is somewhat beneficial: dispersion alone would act to increase the length of the driver to $\sim300$~\si{\micro\metre} after 240~\si{\metre}, which is reduced to $\sim260$~\si{\micro\metre} due to the gradual loss of the beam head.

\begin{figure}
\begin{center}
\includegraphics[width=0.5\textwidth]{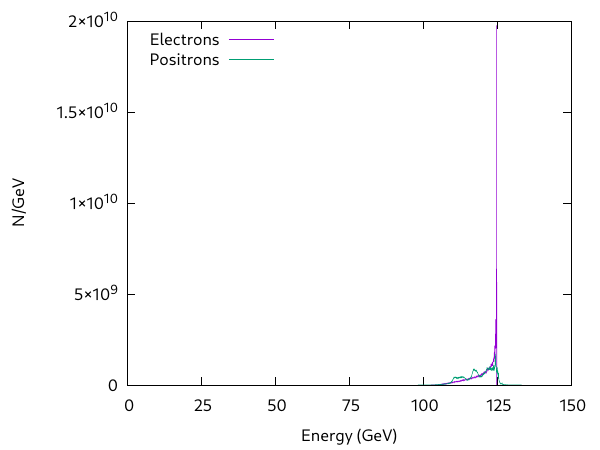}
\caption{{\it Energy spectra of the accelerated electron and positron beams after 200 and 195~\si{\metre}, respectively.  Further optimization of the spectra is possible.}}
\label{fig:spectra} 
\end{center}
\end{figure}

Proof-of-principle simulations for the acceleration of both electrons and positrons were carried out for witness bunches with 10\% of the driver charge and an initial energy of 1~\si{\giga\electronvolt}.  In order to reduce the computational overhead, the azimuthally symmetric quasistatic PIC code LCODE~\cite{pic-lotov-lcode,pic-lcode-manual} was used for these simulations.  Excellent agreement was found with the full 3D simulations for the wakefields excited by the proton drive bunch.  As with the drive bunch, the witness bunches were modelled with a parabolic profile in $z$ and $p_z$.  A bunch length of 105~\si{\micro\metre} was chosen for electrons, and 75~\si{\micro\metre} for positrons, with acceleration to an energy of 125~\si{\giga\electronvolt} in 200~\si{\metre} of plasma.  The spectra of the accelerated bunches are shown in Fig.~\ref{fig:spectra}, with the marked difference between electrons and positrons arising due to the asymmetry of the plasma response.

The electron witness bunch rapidly drives a blowout, with the bulk of the witness sitting inside the cavitated region free of plasma electrons~\cite{olsen-emittance}. For the parameters considered here, ion motion is initially negligible, with the plasma ions providing a linear focusing field which allows the normalized emittance of the witness to be preserved during acceleration.  Adiabatic focussing during acceleration causes the charge density of the witness bunch to increase as $\sqrt{\gamma}$, with the witness bunch only becoming sufficiently dense to perturb the density of lithium ions after several metres of acceleration.  Recent simulation studies have shown that this slow evolution of the focusing field allows the witness bunch to ``self-match'' to the fields acting upon it~\cite{ref:ionmotion}, leading to an emittance growth of only a few percent.  We therefore retain static ions for the proof-of-principle simulations in this work.  In this case, the normalized emittance of the electron bunch is preserved during acceleration.

After acceleration, the witness electron bunch has an RMS energy spread of 3.8\%, with 46\% of the charge within a $\pm$1\% energy range.  The absolute slice energy spread is conserved during acceleration, and so further improvements should be readily achievable through the use of witness bunches with a tailored current profile to control the beam-loading of the wakefields~\cite{meer-beamloading}.  The achievable energy spread is therefore only limited by the witness energy spread at injection and the ability to tailor the witness current profile.

For the positron witness, the focusing comes from the plasma electrons.  As the electron mass is orders of magnitude lower than that of the ions, the focussing field can be strongly nonlinear, even at low witness energies, resulting in significant emittance growth. Furthermore, the plasma electrons execute fast oscillations within the field of the positrons, resulting in a longitudinal variation in the accelerating field~\cite{cao-positronreview}.  Several techniques have been proposed to mitigate these effects.  For beams with moderate charge density, an equilibrium state can be found for a positron witness \cite{hue-positronequilibium}, while controlling the wake structure may allow the acceleration of low-emittance positron beams \cite{silva-ionmotionchannel,diederichs-plasmacolumn,zhou-asymmetricchannel}.  The use of warm plasmas can also act to decrease the nonlinearity of the focussing field \cite{diederichs-warmplasma}.  A recent review of plasma-based positron acceleration can be found in reference~\cite{cao-positronreview}.  We note that none of these techniques are directly applicable to the acceleration scheme proposed here, and that significant development of these concepts is necessary for the realization of proton-driven positron acceleration.  

The positron bunch shows a significantly larger energy spread, as expected due to the plasma response. The asymmetry of the plasma response also results in an unloaded wake with a focusing phase for positrons which is significantly smaller than that for electrons, as seen in Fig.~\ref{fig:phase}c. This constrains the position of the positron bunch, and so the use of a tailored bunch profile would be especially beneficial, providing additional degrees of freedom to optimize the energy spread.  Using different driver configurations for electron and positron acceleration may also allow for further improvements, for example by using a less nonlinear wake for positron acceleration, broadening the focusing phase.  The normalised emittance of the positron bunch grows significantly during acceleration, reaching $\sim150$~\si{\micro\metre}.  As discussed above, no method currently exists to accelerate positrons in this scheme while maintaining their emittance, and new techniques should be investigated as a matter of priority.  The development of novel simulation techniques would be extremely beneficial, as the radius of the positron bunch will be significantly smaller than that of the electrons, and the acceleration length is orders of magnitude longer than in comparable studies.

\section{Evaluation of available Luminosity}

\begin{table}[hbpt]
    \centering
    \begin{tabular}{c|c|cc}
   \hline
    Proton Accelerator Parameter     & Symbol & Unit & Value \\
    \hline
    Proton energy & $E_p$ & \si{\giga\electronvolt} & 400 \\
    Refill Time & $\tau$ & s & 0.2 \\
    Bunch population & $N_p$ & $10^{10}$ & 10 \\
    Number of bunches & $n$ & & 1000 \\
   Longitudinal RMS & $\sigma_z$ & \si{\micro\metre} & 150 \\
  Transverse RMS & $\sigma_{x,y}$ & \si{\micro\metre} & 240 \\
    Normalized transv. emittance & $\epsilon_{T,p}$ & \si{\micro\metre} & $3-75$~\si{\micro\metre}  \\
        Power Usage & P & \si{\mega\watt} & 150 \\
    \hline
    \hline

       Plasma Parameters &Symbol    &  Unit & Value \\
\hline
$e^-$ cell Length & $L_{e^-}$ & \si{\metre} & 240 \\ 
$e^+$ cell Length & $L_{e^+}$ & \si{\metre} & 240 \\ 
density - upstream & $n_p$ & $10^{14}$~cm$^{-3}$ & 3.2 \\
density - downstream & $n_p$  & $10^{14}$~cm$^{-3}$ & 5.1 \\
    \hline
    \hline
       e$^\pm$ Bunch Parameters &Symbol    &  Unit & Value \\
    \hline
Injection Energy & $E_{e,in}$& \si{\giga\electronvolt} & 1 \\
Final Energy & $E_{e}$ & \si{\giga\electronvolt} & 125 \\
Bunch population & $N_{e^{\pm}}$ & $10^{10}$ & 2 \\
Normalized transv. emittance &$\epsilon_{T,e}$ & nm &  $100$\\
Hor. beta fn. &
$\beta^*_x$ & mm &  $13$\\
Ver. beta fn. &
$\beta^*_y$ & mm &  $0.41$\\
Hor. IP size. &
$\sigma_x^*$ & nm & $73$\\
Ver. IP size. &
$\sigma_y^*$ & nm & $13$\\
        \hline
        \hline
    $e^- e^+$ Collider Parameter
    & Symbol &  Unit & Value \\
    \hline
    Center-of-Mass Energy & $E_{\rm cm}$ & \si{\giga\electronvolt} & 250 \\
    Average Collision Rate & $f$ & \si{\kilo\hertz} & 5 \\
    Luminosity & $\mathcal{L}$& cm$^{-2}$s$^{-1}$ & $1.7\times 10^{34}$ \\
    \hline
    \end{tabular}
    \caption{The bunch and collider parameters}
    \label{tab:Parameters}
\end{table}

The simulations described in the previous section show that, with adequate pre-conditioning of the proton bunches, $400$~\si{\giga\electronvolt} proton bunches can be used to accelerate bunches of electrons and positrons to the energies needed for a Higgs Factory.  The results demonstrate that the normalized emittance and the slice energy spread of the electron bunches can be controlled during the acceleration process.  The use of witness bunches with a tailored current profile would allow the accelerating wakefields to be flattened in the vicinity of the witness bunch.  In addition to a narrower final energy spread \cite{lindstrom-energyspread}, this would allow the acceleration of higher-charge witness bunches over longer distances, increasing the efficiency.  With moderate optimization, we expect to be able to accelerate bunches of electrons and positrons with about $20$\% of the population of the proton bunches.  A longer propagation distance will be required in order to reach the same energy with this higher charge, but we note that simulations have demonstrated stable wakefields up to 240~\si{\metre}.  We anticipate that with further optimization a large fraction of the accelerated bunches will be within $1$\% of the nominal energy.  These assumptions will need to be verified in further simulation studies.

The acceleration of positrons in plasma remains more challenging, with simulations showing a significant increase in both the emittance and slice energy spread.  Significant progress has been made for the case of electron-driven wakefield acceleration of positron bunches \cite{hue-positronequilibium,diederichs-plasmacolumn,zhou-asymmetricchannel,diederichs-warmplasma,cao-positronreview}, and further efforts are required to develop new techniques for acceleration in proton-driven schemes.  For the following calculations, we assume that further developments in the field will lead to competitive schemes for positron acceleration, and use the same witness parameters as demonstrated for the electron case.  Alternative methods avoiding the need for positron acceleration are discussed at the end of this section.

The assumed parameters for our approach to a proton-driven PWFA Higgs Factory are summarized in Table~\ref{tab:Parameters}.  The $\beta^\ast$ functions chosen are those foreseen for the ILC~\cite{ref:ILCpars}, and it is assumed that appropriate beam manipulations can be applied in order to achieve the same $\beta^\ast$ to the bunches accelerated in our scheme.  The achievable energy spread depends on the ability to shape the witness beams prior to acceleration and is unknown at this time.  We have not accounted for the impact of this energy spread. The bunch sizes at the IP are then:
$$
\sigma^* = \sqrt{\frac{\epsilon_{T,e}}{\gamma_e}\beta^*}.
$$
and $\gamma_e$ is the relativistic $\gamma$ parameter of the electrons and positrons.  The luminosity is calculated using the formula
$$
\mathcal{L} = f \frac{N_{e^-}N_{e^+}}{4\pi \sigma_x^*\sigma_y^*} 
$$
where $f=n/\tau=5000$~\si{\second^{-1}} is the effective $e^{\pm}$ bunch crossing frequency. Effects such as the hourglass effect are not considered at this point, as we are principally interested in order-of-magnitude estimates.   
While the use of flat witness beams would ease the constraints on the final focussing optics, emittance coupling in plasma wakes can degrade such beams \cite{ref:diederichs_flat}, introducing additional constraints on the acceleration process.

Our estimate of the achievable luminosity is to be understood as a goal, and achieving it will depend critically on, amongst other things, whether the positron emittance and slice energy spread can be preserved during the acceleration process.  However, simulations show that the strong accelerating gradients can be maintained over long distances and the phase control is suitable for acceleration, allowing the energies necessary for a Higgs Factory to be reached with $400$~\si{\giga\electronvolt} proton bunches.

The scheme of proton-driven wakefield acceleration also offers several additional avenues for future study. The achievable acceleration is limited by dispersion of the proton bunch as it loses energy to the plasma wakefield it drives.  Since this wakefield-induced lengthening of the driver scales as $L^2/E_p^3$, with $L$ the acceleration length and $E_p$ the proton driver energy, this suggests that a $t\bar{t}$ collider should be viable with 525~\si{\giga\electronvolt} drive beams.  Similarly, the acceleration of a witness electron bunch to 500~\si{\giga\electronvolt}, as considered in the HALHF scheme~\cite{ref:HALHF}, should be possible with a 1000~\si{\giga\electronvolt} driver, which would avoid the need to develop new positron acceleration schemes.  Another option avoiding positron acceleration would be a gamma-gamma collider~\cite{ref:CilentoDualBDS}.  Although not considered here, we also note that plasma wakefield acceleration is well suited to the acceleration of polarized beams due to the large accelerating gradients~\cite{ref:Jorge_polarized}.  Suitable witness bunches could be generated in a similar way to linear accelerator proposals~\cite{ref:CLIC_CDR}.

\section{Discussion}

Our preliminary investigations indicate that proton bunches, provided they can be produced and accelerated with the parameters described in this article, could be used to accelerate electron and positron bunches to the required energies for a Higgs Factory.  Given our assumptions, the luminosity that could be achieved from rapid-cycling proton synchrotrons installed in existing accelerator tunnels (e.g., CERN SPS tunnel, the FNAL Main Ring tunnel or even the RHIC tunnel at BNL) would be comparable to those of other proposed facilities. Indeed, assuming the availability of pairs of proton bunches with $10^{11}$ protons of energy $E=400$~\si{\giga\electronvolt} per bunch, we anticipate that bunches of $2\times 10^{10}$ electrons and positrons can be accelerated and brought to collision at a center-of-mass energy of $250$~\si{\giga\electronvolt}, as required for Higgs pair-production.  Assuming an effective rate of $5$~\si{\kilo\hertz} and that small emittances will be maintained in the acceleration process, a simple estimate yields a luminosity of $1.7 \times 10^{34}$~cm$^{-2}$s$^{-1}$.  This number is at best an order-of-magnitude estimate of the achievable luminosity, as many technical challenges remain to be resolved.

The key issues for our application will be the development of high-temperature superconducting magnets, allowing for reduced power needs and making high-repetition-rate acceleration energy efficient; the ability to have short, of order $150$~\si{\micro\metre}, proton bunches with tailored emittance profiles; and the development of new techniques to accelerate positrons in plasma.  The emittance profile of the driver could potentially be achieved through the exploitation of plasma instabilities.  The acceleration of positron bunches in plasma while maintaining the beam quality is an area of active research. The ability to recover unused energy from the spent proton bunches will increase the overall energy efficiency.

The development of the required technologies would open the door for much higher-energy lepton colliders.  Additional advantages of developing such a facility would be the availability of the proton bunches as sources of muons for a muon collider and neutrinos for neutrino physics, i.e.\ the spectrum of particle-physics applications is very broad.
 
There are clearly many challenges to the development of a proton-driven plasma wakefield accelerator intended for high-energy collisions.  Given the potential for such an accelerator, these challenges should be taken up and met. 


\section{Acknowledgments}

We would like to thank Ferdinand Willeke, Vladimir Shiltsev, Giovanni Zevi Della Porta, Andrea Latina, Brian Foster and Karl Lindsrøm for several constructive discussions carried out in the course of this study.

\bibliography{Higgs}

\end{document}